\documentclass[aps, prl, 10pt,
amssymb,
amsmath,
superscriptaddress,
tightenlines,
twocolumn,
notitlepage,
] {revtex4}
\usepackage{graphicx}
\usepackage[colorlinks, linkcolor = blue, citecolor = blue, urlcolor=blue, pdfborder={0 0 0 [0 0]},bookmarks=false]{hyperref}
\usepackage{bm}
\usepackage{physics}
\usepackage{amssymb}
\usepackage{amsmath}
\usepackage[dvipsnames]{xcolor}
\usepackage{lipsum}
\usepackage{siunitx}

\newcommand{\be}{\begin{equation}}
\newcommand{\ee}{\end{equation}}

\begin{document}

\title{Wannier functions, minimal model and charge transfer in Pb$_{9}$CuP$_6$O$_{25}$}

\author{Ning Mao}
\affiliation{Max Planck Institute for Chemical Physics of Solids, 01187, Dresden, Germany}

\author{Nikolai Peshcherenko}
\affiliation{Max Planck Institute for Chemical Physics of Solids, 01187, Dresden, Germany}

%\author{Jiangxu Li}
%\affiliation{Department of Physics and Astronomy, University of Tennessee, Knoxville, TN 37996, USA}

%\author{Cristian D. Batista}
%\affiliation{Department of Physics and Astronomy, University of Tennessee, Knoxville, TN 37996, USA}

\author{Yang Zhang}
\email{yangzhang@utk.edu}
\affiliation{Department of Physics and Astronomy, University of Tennessee, Knoxville, TN 37996, USA}
\affiliation{Min H. Kao Department of Electrical Engineering and Computer Science, University of Tennessee, Knoxville, Tennessee 37996, USA}

\begin{abstract}
%The problem of doping Mott insulators and magnetic frustrations is of fundamental importance and long-standing interest in the study of strongly correlated electron systems.
Recent preprints claimed that the copper doped lead apatite Pb$_{9}$CuP$_6$O$_{25}$ (LK99) might be a high-temperature superconductor because of its strong diamagnetism and transport properties. Motivated by the strongly correlated effects that can arise from a triangular lattice of Cu atoms with narrow bandwidth, we calculated the maximally projected Wannier functions from density functional theory simulations, and constructed a minimal two-orbital triangular model with Cu ($3d_{xz},3d_{yz}$) basis, and a four-orbital buckled honeycomb model with Cu ($3d_{xz}$,$3d_{yz}$), O $(2p_x,2p_y)$. Since the Coulomb interaction $U_d$ is much larger than potential energy difference $\Delta$ between Cu and O, charge transfer will occur for hole filling fraction $n_h>1$. We further calculate the interaction parameters, and discuss the possible insulating state and corresponding spin exchange coupling.

\end{abstract}

\maketitle

Despite the multiple ongoing experimental and theoretical studies of Pb$_{9}$CuP$_6$O$_{25}$ (LK99) \cite{lee2023firs,lee2307superconductor,guo2023ferromagnetic,si2023electronic,hou2023observation,kumar2023synthesis,liu2023semiconducting,lai2023first,griffin2023origin,oh2023swave}, it is not clear if this material is a room-temperature superconductor. Different behaviors have been reported from transport measurements (metallic or semiconducting) and magnetic susceptibility (diamagnetic or ferromagnetic). From the theory side, it is found the narrow isolated bands at Fermi level are formed by two Cu $3d$ orbitals, and there are two gapless nodes at $\Gamma$ and $A$ momentum. To study the interesting properties such as band topology and correlated effects, an effective interaction model from material simulations is highly desired.

The Mott-Hubbard theory stands as the main paradigm for the study of strongly correlated  oxides. The Hubbard model captures  the essence of numerous electronic phenomena, including Mott insulators, metal-insulator transitions, metallic ferromagnetism, charge/spin stripe states and unconventional superconductivity. The square lattice Hubbard model was studied intensively in the context of cuprate high-temperature superconductors \cite{lee2006doping}. While the geometric frustrated triangular lattice Hubbard models are relatively rare in nature, notable examples are artificial semiconductor moir\'e systems \cite{regan2020mott,tang2019wse2,PhysRevLett.121.026402,PhysRevB.102.201115,li2021continuous,https://doi.org/10.48550/arxiv.2202.02055} and several superconducting systems: Na$_x \mathrm{CoO}_2 \cdot y \mathrm{H}_2 \mathrm{O}$ \cite{takada2003superconductivity,wang2004doped}, Sn at Silicon 111 surface \cite{ming2023evidence}, and most recently the debated LK99 Pb$_{9}$CuP$_6$O$_{25}$ \cite{lee2023firs,lee2307superconductor}.

In Pb$_{9}$CuP$_6$O$_{25}$, the doped Cu atoms form a triangular lattice with a large parameter $a \simeq 1$~nm in the Cu-O plane.
Compared to other correlated materials, Pb$_{9}$CuP$_6$O$_{25}$  has a largely reduced kinetic energy, which leads a bandwidth $\sim 140$ meV, as predicted by density functional theory (DFT) calculations~\cite{lai2023first,griffin2023origin}. Since the bare onsite Coulomb repulsion between Cu~$3d$ orbitals is or the order of 10 eV  (strong coupling limit with $U_d>>t$), the  Cu-O hopping amplitude and the charge transfer gap should be considered when constructing the effective low-energy-model of this material. Having a full Cu-O four band Hubbard model, one can then derive an effective $t-J$ model on triangular lattice, if the system is a charge transfer insulator for hole filling fraction $n_h=1$ and the low energy states is described by well-seperated Zhang-Rice single states.

In this work, we study the electronic bands near Fermi level with DFT and derive the minimal two-orbital triangular lattice model and four-orbital honeycomb lattice model for Pb$_{9}$CuP$_6$O$_{25}$. The distinct arrangement of Cu-O bonds leads to the Cu $3d$ orbital splitting into a doublet ($3d_{yz}$,$3d_{xz}$), singlet $3d_{z^2}$, doublet ($3d_{xy}$,$3d_{x^2-y^2}$) structure. We project the well isolated bands near Fermi level to the maximally projected Wannier functions using Cu basis ($3d_{yz}$,$3d_{xz}$) and the combined Cu ($3d_{yz}$,$3d_{xz}$), O $(2p_x,2p_y)$. Starting from  a tight-binding model constructed with Wannier orbitals, we keep only the nearest neighbor hoping in Cu-O plane and out-of-plane, which matches well with the electronic structures from DFT. We note that the two models have the same topological features, the $C_3$ and time reversal symmetry protected spinless double Weyl nodes at $\Gamma$ and $A$ without spin-orbital coupling, and a pair of chiral surface states arises when Weyl nodes are gaped by spin-orbit coupling.

%Following that,
We then proceed to calculate the Coulomb interaction and spin exchange coupling from the projected Wannier functions and construct the four-orbital honeycomb Hubbard model. Given the small potential energy difference ($\Delta\sim$ 0.2 eV$<<U_d$) between Cu~$3d$ and O~$2p$, charge transfer from Cu to O will happen for hole filling fraction $n_h>1$. Consequently, the dominant effective spin-spin interaction the antiferromagnetic super-exchange mediated by the 2$p$-orbitals of the corner O atoms. We note the large Cu to O hopping strength $t_{pd} \simeq$ 0.1 eV might lead to a metallic state or charge transfer insulator at $n_h=1$.

%With the effective Hubbard model, we analyze the magnetic configurations derived from a unique filling factor: three electrons over four orbitals considering spin degrees of freedom.
%On the superconductivity side, we believe the kinetic energy is too small for the suspected high Tc ($\sim 110$ K) from transport experiments \cite{hou2023observation}, and we may need different lattice structures or new type of pairing mechanism.
\begin{figure}[!h]
\includegraphics[width= 0.5\textwidth]{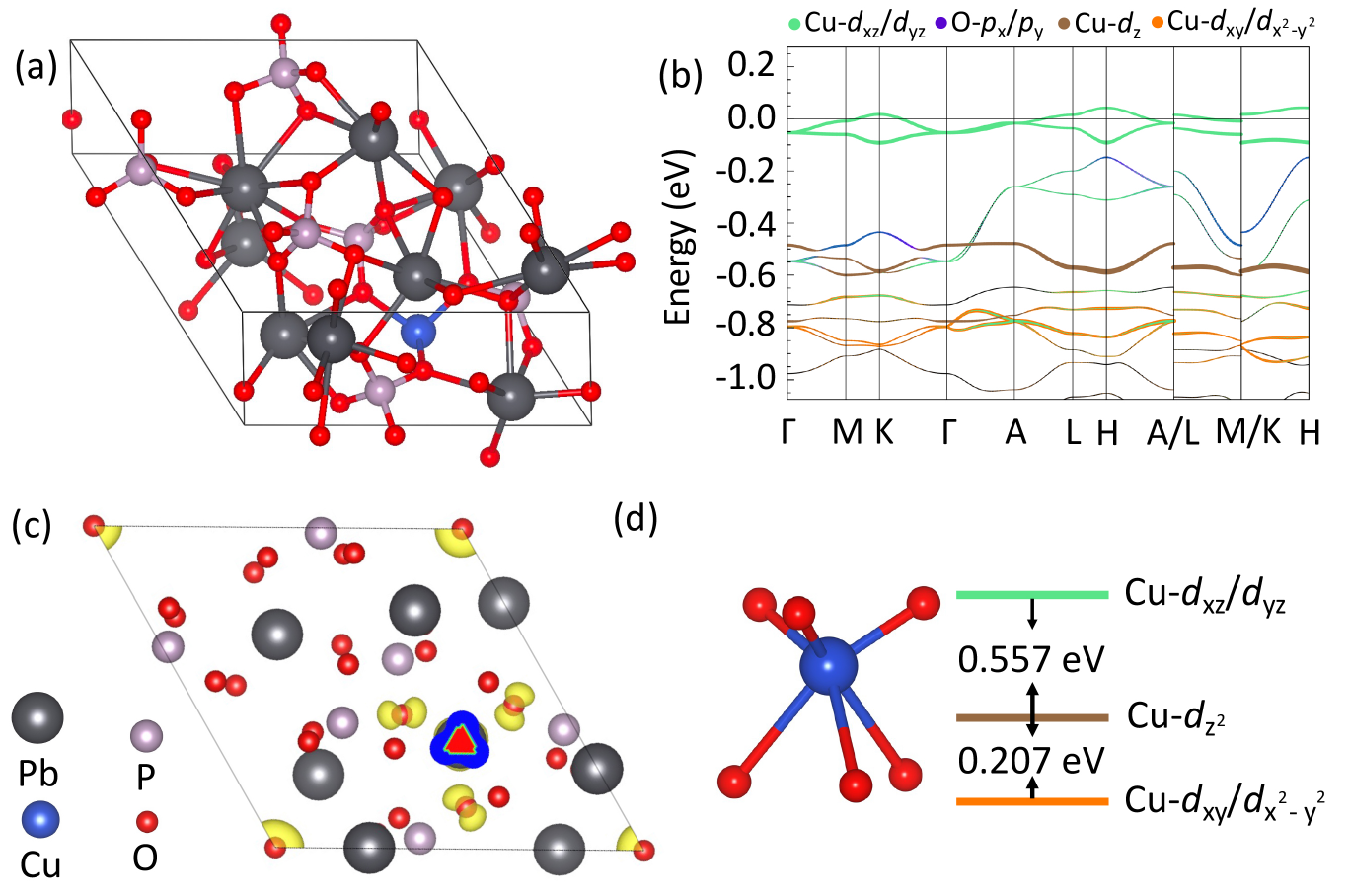}
\caption{
(a) Lattice structure of Pb$_{9}$CuP$_6$O$_{25}$;
(b) Electronic band structure without spin-orbit coupling, Orbital weights of Cu and corner O atoms are plotted with different colors;
(c) Charge density distribution of two bands near Fermi level, with $<5\%$ weights on O atoms;
(d) Six Cu-O bonds with three fold rotation symmetry, and the crystal field splitting.
} \label{fig:dft}
\end{figure}

\begin{figure}[ht]
\includegraphics[width= 0.5\textwidth]{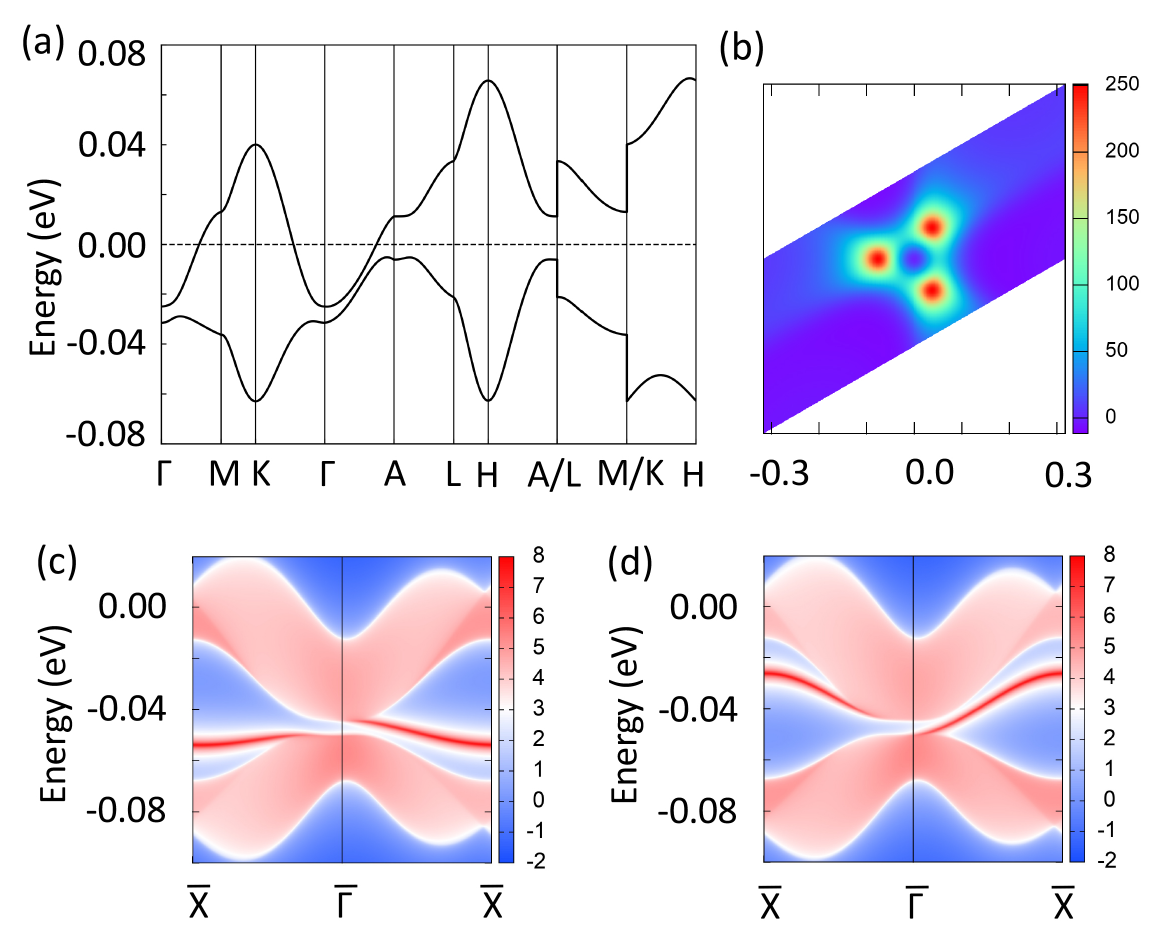}
\caption{
(a) Electronic band structure with spin-orbit coupling, and the double Weyl nodes are gaped out;
(b) The reciprocal-space distribution of Berry curvature within the bandgap;
(c) Surface state of upper surface;
(d) Surface state of lower surface.
} \label{fig:tp}
\end{figure}

\textbf{First principle simulation and Maximally Projected Wannier Functions.}
We perform the density functional theory simulation with Full-Potential-Local-Orbital code, FPLO \cite{koepernik1999full}. The relaxed lattice structure is taken from Ref. \cite{lai2023first}
(lattice constants are $\SI{9.88}{\angstrom}$, $\SI{9.88}{\angstrom}$, and $\SI{7.40}{\angstrom}$ ), and Pb(1) is replaced by Cu atom in Fig. \ref{fig:dft}(a). Without spin-orbital coupling, the calculated bandwidth is around 140 meV, similar to other first principle simulations\cite{lai2023first,griffin2023origin}. In Fig. \ref{fig:dft}(b,c), the relevant bands near Fermi level are formed from doped Cu $3d$ orbitals with $<\%5$ contribution from O atoms. The conduction band minimum are Pb $6p$ orbitals and O $2p$ orbitals, with a large separation around 4 eV to Cu $3d$ states. The degeneracy at $\Gamma$ and $A$ near Fermi level are protected by spinless $C_3$ symmetry, and will be gaped after considering spin-orbital coupling.

Due to three fold rotation symmetry of Cu-O bonds, the Cu $3d$ orbitals split into three groups: doublet ($3d_{xz}$,$3d_{yz}$), singlet $3d_{z^2}$, and doublet ($3d_{xy}$,$3d_{x^2-y^2}$). From the orbital weights analysis, we find that two bands at Fermi level are dominantly formed from doublet $3d_{yz}$ and $3d_{xz}$. The crystal field splitting of the Cu $3d$ orbitals are shown in Fig. \ref{fig:dft}(d), with 0.557 eV and 0.207 eV splitting in the ($3d_{xz}$, $3d_{yz}$), $3d_{z^2}$, and ($3d_{xy}$, $3d_{x^2-y^2}$) order.

After the band structure calculations, we project the Bloch wavefunctions into two different basis sets: (1) two orbitals as Cu ($3d_{xz}$,$3d_{yz}$), and (2) four orbitals as Cu ($3d_{xz}$,$3d_{yz}$), O $(2p_x,2p_y)$. Unlike the maximum localization scheme \cite{marzari2012maximally}, our gauge fixing matrix for transforming Bloch wavefunctions into Wannier functions is obtained by picking linear combinations of wave functions of a Hilbert subspace which have maximal projection (band weight) for chosen local orbitals basis \cite{koepernik2023symmetry}. This approach retains the symmetry properties of Wannier functions, and yields a very high degree of localization due to the nature of the local orbitals.

\begin{figure}[!h]
\includegraphics[width= 0.5\textwidth]{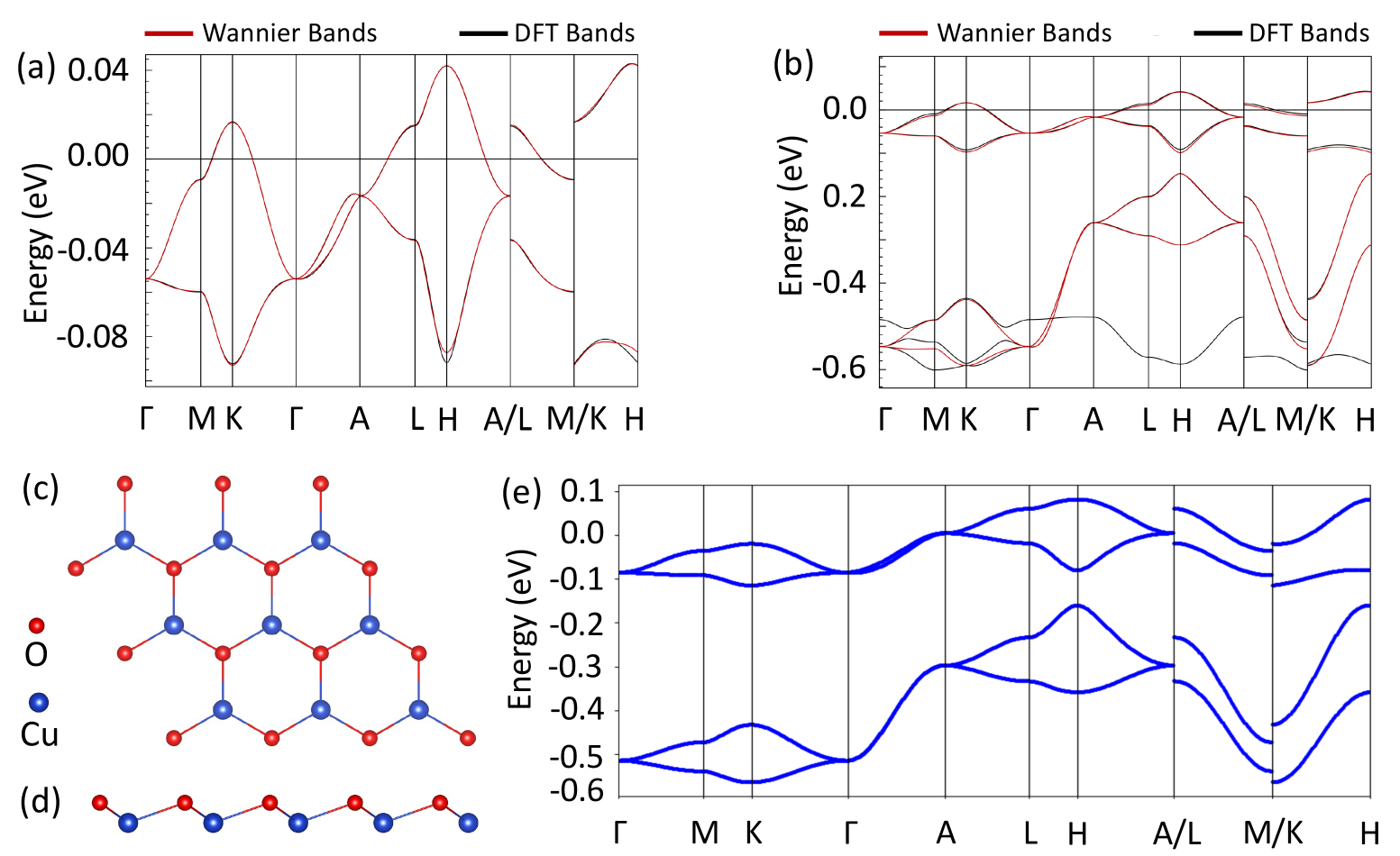}
\caption{
(a) Comparison of DFT bands with those obtained by two-orbital Wannier interpolation;
(b) Comparison of DFT bands with those obtained by four-orbital Wannier interpolation;
(c) Top and (d) side view of buckled honeycomb lattice made of Cu and O atoms;
(e) Electronic band structure from four-orbital tight-binding model.
} \label{fig:wan}
\end{figure}

To increase the fitting quality, we apply the automatic Wannierization scheme \cite{zhang2021different} and obtain the ideal Wannier bands as shown in Fig. \ref{fig:wan}(a) - \ref{fig:wan}(b), with mean errors less than 5 meV. Considering spin degree of freedom, the four bands near Fermi level are filled with three electrons after integrating the density of states. We further plot the Wannier functions in real space and find the $3d$ Wannier functions are localized around Cu site with a spread less than $\SI{5.0}{\angstrom}$.

\textbf{Tight binding model of Cu $3d$ orbitals and topological bands.} Starting from the Wannier hopping parameters, we write the effective tight-binding model with only nearest neighbour hopping along Cu-O plane and out-of-plane directions. The second nearest neighbour in-plane hoppings are listed in the supplementary materials, and further long range hoppings are less than 1 meV.
\begin{align}
	H_c =&  \sum_{ im}  e_{i} c_{im }^{\dagger} c_{im} + \sum_{\langle i, j\rangle,mm'  }  T_{l}^z c_{im }^{\dagger} c_{jm'} \\
 +& \sum_{\langle i, j\rangle,mm'}  T_{l}^{xy} c_{im }^{\dagger} c_{jm'} +h . c .,
\end{align}
where $c_{i m\sigma}^{\dagger}(c_{i m\sigma})$ is the Fermion creation (annihilation) operator for orbital $m$ on site $i$, and the angular brackets in $\langle ... \rangle$ restrict the sums to nearest neighbor hopping.

Under the basis of $3d_{xz}$- and $3d_{yz}$-orbitals of the Cu atoms, the hopping matrices $T^z_{l}$ and $T^{xy}_{l}$ are represented as 2 $\times$ 2 matrices, and $l$ denote for the Cu-Cu bond index as shown in Fig. \ref{fig:tb}.
\begin{align}
	T_{1}^z=\left(\begin{array}{cc}
		t_{1}^z & -t_{2}^z \\
		t_{2}^z & t_{1}^z
	\end{array}\right)
\end{align}
\begin{align}
	T_{1}^{xy}=\left(\begin{array}{cc}
		t_{\sigma} & t_{1}+t_{2} \\
		t_{1}-t_{2} & t_{\pi}
	\end{array}\right)
\end{align}
Here, the parameters $t_1^z$ and $t_2^z$ correspond to the out-of-plane interlayer hopping strengths. In contrast, $t_{\sigma}$, $t_{\pi}$, $t_1$, and $t_2$ denote for the in-plane intralayer hopping strengths. Notably, $t_{\sigma}$ and $t_{\pi}$ stem from the unique overlaps of the $d_{xz}$ and $d_{yz}$ orbitals. Concurrently, $t_1$ and $t_2$ originate from the bond integrals of $\sigma$ and $\pi$ bonds, respectively. From the projected Wannier Hamiltonian, the parameters are extracted as: $e_1$ = -29.6 meV, $e_2$ = -29.6 meV, $t_1^z$ = -6.16 meV, $t_2^z$ =  -0.32 meV, $t_{\sigma}$ = -4.26 meV, $t_{\pi}$ = 2.38 meV, $t_1$ = -6.09 meV, and $t_2$ = 10.99 meV.

It's crucial to note that a non-zero $t_2$ breaks the two-fold rotational symmetries for the in-plane direction, resulting in a transition of the point group from $D_3$ to $C_3$. Therefore, the Hamiltonian solely retains the three-fold rotational symmetries. The hopping matrix for the other nearest neighbor bonds can be deduced by $C_3$ rotation operations:
\begin{align}
	C_{3}=\left(\begin{array}{cc}
		-\frac{1}{2} & \frac{\sqrt{3}}{2} \\
		-\frac{\sqrt{3}}{2} & -\frac{1}{2}
	\end{array}\right),
	C_3^{-1}  = \begin{pmatrix}
		-\frac{1}{2} & -\frac{\sqrt{3}}{2} \\
		\frac{\sqrt{3}}{2} & -\frac{1}{2}
	\end{pmatrix}
\end{align}
In crystal momentum space, the $C_3$ symmetry can only protect degeneracies at the invariant lines or points of the Brillouin Zone, characterized by $C_3 \mathbf{k} \rightarrow \mathbf{k}$. Within this invariant space, the Hamiltonian commutes with the symmetry and can be block diagonalized into three sectors, each having distinct eigenvalues. Along the $C_3$-invariant line ($\Gamma A$), each bands are marked by specific $C_3$ symmetry eigenvalues, $e^{ix 2\pi/3}$ ($x$ = 0, 1, 2). Constrained by the time reversal symmetry $\mathcal{T}$, one band denoted by $e^{i 2 \pi/3}$ will be degenerated with another band denoted by $e^{i 4\pi /3}$. The degeneracy is robustly protected by $C_3$ and $\mathcal{T}$ symmetries. As a consequence, the only points encapsulating both the $C_3$ and $\mathcal{T}$ symmetries within the three-dimensional Brillouin zone are the $\Gamma$ and $A$ points. At these points, the $d_{xz}$ and $d_{yz}$ (as well as $d_{xy}$ and $d_{x^2-y^2}$) orbitals will embody conjugate $C_3$ symmetry eigenvalues, forming into a twofold-degenerate representation, E. It has to be mentioned that any $C_3$ allowed terms cannot break the degeneracy. Instead, they merely adjust the energy of the degenerate point, leading to the presence of $C_3$-protected crossing point. Conversely, the $d_{z^2}$ orbital will give rise to the one-dimensional representation, A. When considering spin-orbit coupling, due to the breaking of $\mathcal{T}$ symmetry with magnetic moment, the twofold degenerate points will manifest a band gap.

As depicted in Fig. \ref{fig:tb}(b), the band structure reveals two parabolic crossings at the $\Gamma$ and $A$ points. Notably, the shapes of these crossings are similar to each other represented in the Fig. \ref{fig:tb}(e) and \ref{fig:tb}(f). The Fermi surface, illustrated in Fig. \ref{fig:tp}(d), highlights the semi-metallic characteristics in the single particle band structure.

Interestingly, the degenerated states are split out when spin-orbit coupling is taken into account, opening the band gap for $\Gamma$ and $A$ points, as shown in Fig. \ref{fig:tp}(a). As we expected, the spin-orbit coupling split the degeneracy of $\Gamma$ and $A$ points in the Brillouin zone and give rise to the accumulation of the Berry curvature as shown in Fig. \ref{fig:tp}(b), turning the system to be a topologically nontrivial one. To validate the topological property, we investigate the emergence of surface states in our system, which are the hallmarks of nontrivial phases. As shown in Fig. \ref{fig:tp}(c) and \ref{fig:tp}(d), a pair of chiral surface states arise around the Fermi level, exhibiting the property of Chern metal.

\textbf{Four-orbital tight binding model on honeycomb lattice.} As discussed previously, the charge transfer between Cu and O is important when considering the doping induced correlated states beyond $n_h>1$. Here we construct a four-band tight-binding Hamiltonian composed by $2p_x$ and $2p_y$ orbital of O atom, as well as $3d_{xz}$ and $3d_{yz}$ orbital of Cu atoms:
\begin{align}
	H =\left(\begin{array}{cc}
		H_{D} & H_{DP} \\
		H_{PD} & H_{P}
	\end{array}\right)
\end{align}
\begin{align}
	H_D &=  \sum_{ im}  d_{i} c_{im }^{\dagger} c_{im} + \sum_{\langle i, j\rangle,mm'  }  D_{l}^z c_{im }^{\dagger} c_{jm'} \\
 &+ \sum_{\langle i, j\rangle,mm'}  D_{l}^{xy} c_{im }^{\dagger} c_{jm'} +h . c .
\end{align}
\begin{align}
	H_P &=  \sum_{ im}  p_{i} c_{im }^{\dagger} c_{im} + \sum_{\langle i, j\rangle,mm'  }  P_{l}^z c_{im }^{\dagger} c_{jm'} \\
 &+ \sum_{\langle i, j\rangle,mm'}  P_{l}^{xy} c_{im }^{\dagger} c_{jm'} +h . c .
\end{align}
\begin{align}
	H_{DP} =  \sum_{\langle i, j\rangle,mm'}  DP_{l}^{xy} c_{im }^{\dagger} c_{jm'} +h . c .
\end{align}
Here $H_D$ stands for hopping between Cu $3d$ orbitals, $H_P$ stands for hopping between O $2p$ orbitals, and $H_{DP}$ is the hybridization between Cu $3d$ and O $2p$ orbitals. The onsite energy from Wannier tight binding model is given as $d_1 = d_2$ = -109.5 meV, $p_1 = p_2$ = -313.0 meV. The 2$\times$2 hopping matrix $D_{l}^z$, $D_{l}^{xy}$, $P_{l}^z$, $P_{l}^{xy}$, and $DP_{l}^{xy}$ are hopping process can be written as the following form (in the unit of meV):
\begin{align}
	D_{1}^z=\left(\begin{array}{cc}
		  -3.8 &  8.6 \\
		-8.6 & -3.8
	\end{array}\right), %%% exp( -1j *kz)
	D_{1}^{xy}=\left(\begin{array}{cc}
		2.5 & -4.0 \\
		-5.9 & -7.6
	\end{array}\right)
\end{align}
\begin{align}
	D_{2}^{xy}=\left(\begin{array}{cc}
		-0.8 &  5.9 \\
		  7.8& -4.3
	\end{array}\right),
	D_{3}^{xy}=\left(\begin{array}{cc}
		-9.3 & -2.8 \\
		  -0.9 &  4.2
	\end{array}\right).
\end{align}
\begin{align}
	P_{1}^z=\left(\begin{array}{cc}
		-68.2 &  -1.6 \\
		 1.6 &   -68.2  %%% exp(-1j kz)
	\end{array}\right),
	P_{1}^{xy}=\left(\begin{array}{cc}
		-4.2 & -3.9\\
		  1.2 &  1.7
	\end{array}\right)
\end{align}
\begin{align}
	P_{2}^{xy}=\left(\begin{array}{cc}
		1.4 & 0.6 \\
		  -4.4 &  -3.9
	\end{array}\right),
	P_{3}^{xy}=\left(\begin{array}{cc}
		-0.9 & 5.8 \\
		  0.7 &  -1.6
	\end{array}\right).
\end{align}
\begin{align}
	DP_{1}^{xy}=\left(\begin{array}{cc}
		58.5    & 48.6  \\
		    -45.0  & -85
	\end{array}\right),
	DP_{2}^{xy}=\left(\begin{array}{cc}
		  -50.4   & 108.2   \\
		 14.3  &   24.1
	\end{array}\right)  %%% exp( 1j kx)
 \label{transition_matrices}
\end{align}
\begin{align}
	DP_{3}^{xy}=\left(\begin{array}{cc}
		 -47.5   &  -15.9  \\
		 -109.8     &  21.3
	\end{array}\right),  %%% exp (-1j ky)
\end{align}

\textbf{Hubbard and exchange interactions} Adding the interacting terms, we construct the four-band honeycomb lattice Hubbard model with nearest-neighbour hopping at electron filling $n=3$ ($n$ is the number of electrons per unit cell, and full filling is $n_0=4$):
\begin{align}
\begin{split}
H = &-\sum_{\langle i, j\rangle,mm'\sigma}\left(t_{ijmm'}c_{im\sigma}^{\dagger} c_{jm'\sigma}+h . c .\right) + \Delta \sum_{i \in O} n_i\\
&+\sum_{im} U_{m}n_{im \uparrow} n_{im \downarrow}+\frac{1}{2}U_{12} \sum_{i,m\neq m'} n_{im} n_{im'} \\
&-\frac{1}{2}J_{12} \sum_{i,m\neq m'} \mathbf{s}_{im} \mathbf{s}_{im'}
-\frac{1}{2}J^z_{12} \sum_{\langle i, j\rangle_z,m} \mathbf{s}_{im} \mathbf{s}_{im}
\end{split}
\label{Hubbard_ham}
\end{align}
where $c_{i \sigma}^{\dagger}(c_{i \sigma})$ is the Fermion creation (annihilation) operator for spin $\sigma$ on site $i$, $n_{i \sigma}=c_{i \sigma}^{\dagger} c_{i \sigma}$ is the number operator, $\Delta=203$ meV is the energy difference between Cu and O, $t_{ijmm'}$ is the hopping between $m$ orbital at site $i$ to $m'$ orbital at site $j$, $U_m$ is the onsite Coulomb repulsion for orbital $m$, $U_{12}$ is the onsite Coulomb repulsion between two Cu orbital, and $J_{12}$ is the onsite exchange (ferromagnetic Hund's coupling) two Cu orbital.

We calculate the density-density interaction and inter orbital spin exchange interaction between real space Wannier functions $|\mathbf{R}, m\rangle$ and $|\mathbf{R}', m'\rangle$.
\begin{equation}
U_{\mathbf{R}^{\prime} m^{\prime}, \mathbf{R} m}=\sum \iint d \mathbf{r} d \mathbf{r}^{\prime}\left|\psi_{\mathbf{R}^{\prime} m^{\prime}}\left(\mathbf{r}^{\prime}\right)\right|^2 \frac{e^2}{\epsilon\left|\mathbf{r}-\mathbf{r}^{\prime}\right|}\left|\psi_{\mathbf{R} m}(\mathbf{r})\right|^2,
\end{equation}

\begin{equation}
\begin{aligned}
J_{\mathbf{R}^{\prime} m^{\prime}, \mathbf{R} m}= & \sum\iint d \mathbf{r} d \mathbf{r}^{\prime} \psi^{*}_{\mathbf{R}^{\prime} m^{\prime}}\left(\mathbf{r}^{\prime}\right) \psi_{\mathbf{R} m}^{*}(\mathbf{r})  \\
& \frac{e^2}{\epsilon\left|\mathbf{r}-\mathbf{r}^{\prime}\right|} \psi_{\mathbf{R}^{\prime} m^{\prime}}(\mathbf{r}) \psi_{\mathbf{R} m}\left(\mathbf{r}^{\prime}\right),
\end{aligned}
\end{equation}
And the bare interaction strengths are $\epsilon U_d=\epsilon U_1=\epsilon U_2=13.6$ eV, $\epsilon U_{12}=12.8$ eV, $\epsilon J_{12}=0.18$ eV, and on-site repulsion for O $2p_x$ and $2p_y$ orbitals are $\epsilon U_p=\epsilon U_3=\epsilon U_4=6$ eV. Non-local interactions are parametrically smaller than onsite $U$ and we estimate the nearest neighbour $\epsilon V_{11},\epsilon V_{22}\sim 1.3$ eV. We compute the dielectric constant using density functional perturbation theory \cite{souza2002first}, and obtained $\epsilon=6$. The Coulomb repulsion $U_d$ is still much larger than potential difference $\Delta$ between Cu $3d$ and O $2p$ orbitals. At hole filling $n_h>1$, the Cu atoms are half filled with one holes per site, and charge transfer to O $2p$ orbitals will happen.

However, here the nearest neighbor hopping between Cu and O $t_{pd}\sim 100$ meV is comparable to $\Delta$, further numerical simulations using spinfull lattice model need to be done to determine whether it is a charge transfer insulator or a metallic state and the critical $\Delta/t$ for metal-insulator transition.

\textbf{Kinetic exchange and super-exchange.} We now consider the effective spin exchange in Pb$_{9}$CuP$_6$O$_{25}$. We consider the original Hubbard model at $n_h=1$ when only Cu sites are filled with one hole per unit cell, the effective spin exchange Hamiltonian between Cu is written as:
\begin{align}
\begin{split}
H_S = &\frac{1}{2}\sum_{\langle i, j\rangle,m=1,2} J^K_{ij,m}s_{im} s_{jm }
%&+\frac{1}{2}J^z_{12} \sum_{i,m\neq m'} s_{im} s_{im'} \\
%+\frac{1}{2}J^t_{12} \sum_{i,m\neq m'} s_{im} s_{im'}
+\frac{1}{2}\sum_{\langle i, j\rangle,m=1,2} J^S_{ij,m}s_{im } s_{jm }
\end{split}
\end{align}
The onsite Hund coupling $J_{12}$ is large, but not relevant for the magnetic properties at $n_h=1$. Here $J^K_{ij,m}$ denotes the Kinetic exchange coupling between nearest neighbor Cu atoms, and $J^S_{ij,m}$ denotes the super-exchange coupling between nearest neighbor Cu atoms mediated by O atoms in Fig \ref{fig:wan}. Given small hopping between Cu sites and large onsite repulsion, the kinetic antiferromagnetic exchange $\frac{4t^2_{dd}}{U_d}\sim 0.1$ meV is vanishing small. While the super-exchange term mediated by empty corner O atom is relatively larger, given by $J_{12}=4\frac{t_{pd}^4}{\Delta^2}(\frac{1}{U_d}+\frac{1}{U_p+2\Delta})\sim 10$ meV from perturbation theory (we use single hopping parameter for simplicity, detailed derivation can be found at SM).

To summarize, the main finding of our work is that the low energy electronic structures in Pb$_{9}$CuP$_6$O$_{25}$ (LK99) can be well described by minimal two-orbital triangular lattice model, and a four-orbital model at honeycomb lattice. For the narrow bands near Fermi level, we observed two double Weyl nodes without spin-orbital coupling, and chiral surface states when two double Weyl nodes are gaped by a small spin-orbital coupling ($\lambda\sim 3$ meV). After comparing the Coulomb repulsion with potential difference between Cu and O, we conclude charge transfer to O $2p$ orbitals will happen at $n_h>1$, and the dominant spin exchange will be antiferromagnetic super-exchange mediated by corner O atoms in Cu-O plane. As the hopping strength is comparable to $\Delta$, further numerical investigations is needed to confirm whether it is metallic or a charge transfer insulator in Honeycomb lattice \cite{zaanen1985band,zhang2020moire}.

In the case of charge-transfer insulators, at temperatures below the charge gap, double occupancy is strongly suppressed by the on-site repulsion $U_d$. And the low-energy physics is described by the $t$-$J$ model \cite{spalek2007tj,lee2006doping} with hopping $t$ and antiferromagnetic super-exchange interaction $J$ between Zhang-Rice-like singlet \cite{zhang1988effective} if the gap from singlet state to excited states is large. The Zhang-Rice-like singlet is in general the hybridized spin-singlet state of $3d$ and $2p$ orbitals in Cu site and corner O site, and the effective $t$ and $J$ depend on the charge transfer gap $\Delta$.

\section*{Acknowledgments}
Y. Z. thanks Zhehao Dai, Yan Sun, JiangXu Li and Shu Zhang for helpful discussions, and especially Cristian D. Batista for the insightful discussions on four band model and valuable comments. Y. Z. was supported by the start up fund at University of Tennessee Knoxville.

%\vfill
%\clearpage

\bibliography{ref}

\clearpage
\pagebreak
\onecolumngrid
\begin{center}
\textbf{\large Supplemental Materials}
\end{center}
\setcounter{figure}{0}
\renewcommand{\thefigure}{S\arabic{figure}}
\setcounter{equation}{0}
\renewcommand{\theequation}{S\arabic{equation}}
\setcounter{table}{0}
\renewcommand{\thetable}{S\arabic{table}}

\subsection{Tight-binding model with spin-orbit coupling}

The hopping matrix of bond 3 ($T_{3}^{xy}$) and 5 ($T_{5}^{xy}$) could be obtained by $C_3$ rotation from $T_{1}^{xy}$.
\begin{align}
	4T_{3}^{xy}=\left(\begin{array}{cc}
		t_{\sigma}+3 t_{\pi} - 2 \sqrt3 t_1 & \sqrt{3} t_{\sigma}- \sqrt{3} t_{\pi}  -2 t_1+4 t_2)\\
		\sqrt{3} t_{\sigma}- \sqrt{3} t_{\pi} -2 t_1 - 4t_2 &    3 t_{\sigma}+t_{\pi} + 2\sqrt3 t_1
	\end{array}\right)
\end{align}
\begin{align}
	4T_{5}^{xy}=\left(\begin{array}{cc}
		t_{\sigma}+3 t_{\pi}  +  2 \sqrt3 t_1 & -\sqrt{3} t_{\sigma}  + \sqrt{3} t_{\pi} - 2t_1+4t_2 \\
		-\sqrt{3}  t_{\sigma} + \sqrt{3} t_{\pi} - 2 t_1-4 t_2&  3 t_{\sigma}+t_{\pi} - 2\sqrt3 t_1
	\end{array}\right).
\end{align}
Furthermore, the term of spin-orbit coupling can be expresented as:
\begin{align}
    H_{soc} =\left(\begin{array}{cc}
		0 &  -i \lambda \\
		i \lambda & 0
	\end{array}\right)
\end{align}
Here, $\lambda$ = 3 meV denotes for the strength of spin-orbit coupling, which is obtained from the Wannier projection with spin-orbit coupling.

\begin{figure}[!h]
\includegraphics[width= 0.5\textwidth]{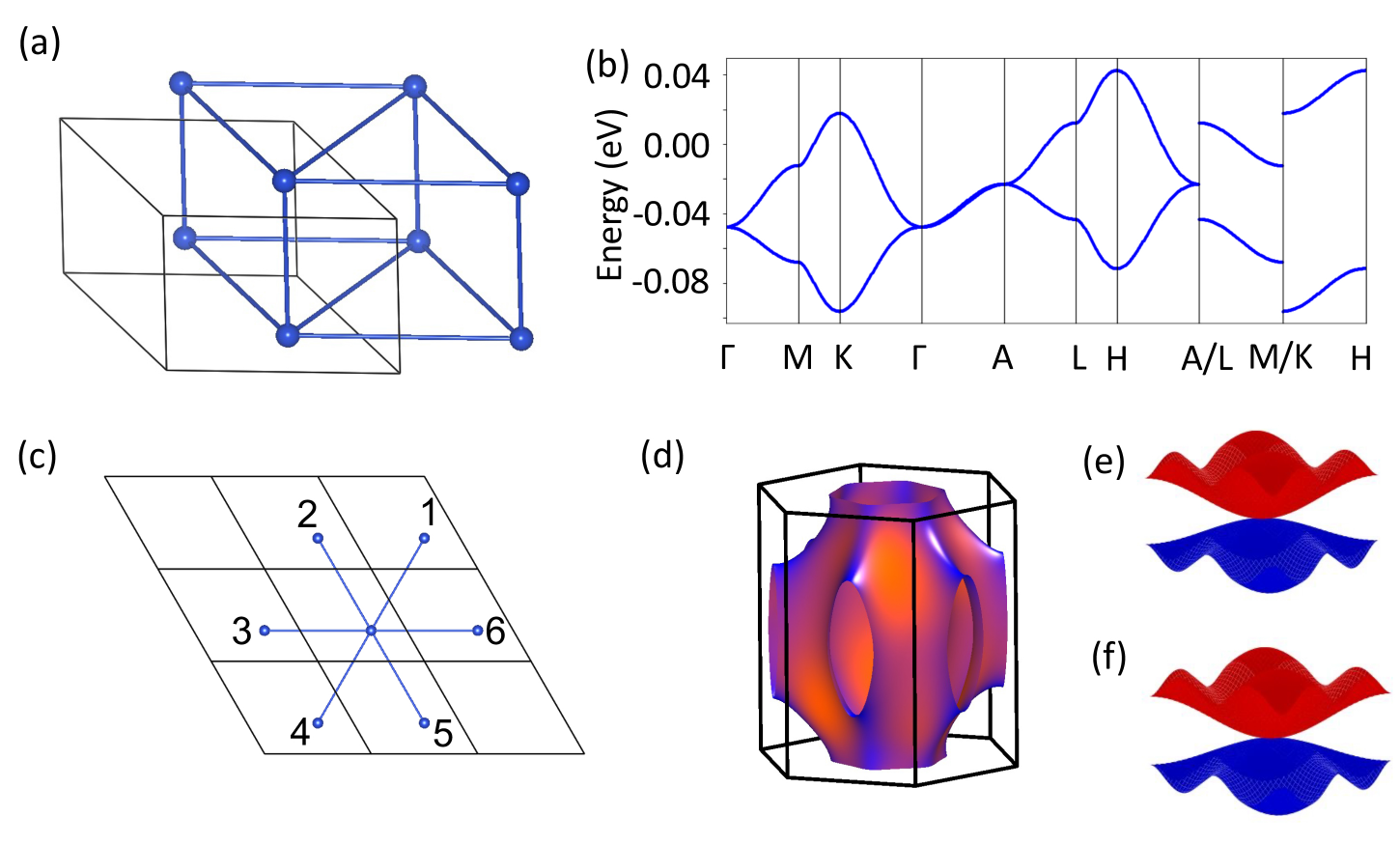}
\caption{
(a) Lattice structure of Cu atoms;
(b) Electronic band structure from simplified tight-binding model;
(c) In-plane hopping path;
(d) Fermi surface plot at $E_f=0$ eV from Wannier tight-binding model;
(e) Three-dimensional band structure with $k_z$ = $\pi$;
(f) Three-dimensional band structure with $k_z$ = $ 0 $.
} \label{fig:tb}
\end{figure}

\subsection{Tight-binding model with longer range hopping}
\begin{figure}[!h]
	\includegraphics[width= 0.6\columnwidth]{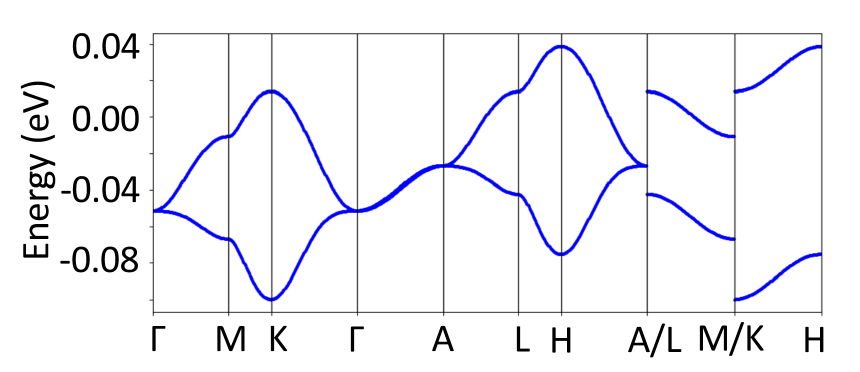}
	\caption{ Electronic band structure with longer range hopping. }
 \label{longer_range}
\end{figure}

We consider a longer range hopping by including the  next-nearest neighbour hopping along in-plane directions:
\begin{align}
	H = \sum_{\langle \langle i, j\rangle\rangle,mm'}  T_{ij}^{xy} c_{im }^{\dagger} c_{m'j } +h . c . ,
\end{align}
where the angular brackets in $\langle ... \rangle$ restrict the sums to next-nearest neighbour hopping, and the three hopping matrix could be represented by:
\begin{align}
	T_{1}^{xy}=\left(\begin{array}{cc}
		n_{1} & n_{2} \\
		n_{3} & n_{4}
	\end{array}\right)
\end{align}
\begin{align}
	4T_{3}^{xy}=\left(\begin{array}{cc}
		 n_{1} - \sqrt3 n_2- \sqrt3 n_3 + 3 n_4 &  \sqrt3 n_1 + n_{2} - 3n_3 - \sqrt3 n_4 \\
		 \sqrt3 n_1 - 3 n_{2} + n_3 - \sqrt3 n_4   & 3 n_1 +\sqrt3 n_2+\sqrt3 n_3 + n_4
	\end{array}\right)
\end{align}
\begin{align}
	4T_{5}^{xy}=\left(\begin{array}{cc}
		 n_{1} + \sqrt3 n_2 + \sqrt3 n_3 + 3 n_4 &  -\sqrt3 n_1 + n_{2} - 3n_3 + \sqrt3 n_4 \\
		 -\sqrt3 n_1 - 3 n_{2} + n_3 + \sqrt3 n_4   & 3 n_1 -\sqrt3 n_2 - \sqrt3 n_3 + n_4
	\end{array}\right) .
\end{align}
And the hopping strengths are $n_1$ = -0.56 meV, $n_2$ = -0.53 meV, $n_3$ = -0.31 meV, and $n_4$ = -0.71 meV. Relative to the nearest neighbour in-plane hopping, the next-nearest neighbour in-plane hopping is approximately an order of magnitude weaker. Therefore, the band structure doesn't change significantly when considering longer range hopping as shown in Fig. \ref{longer_range}.

\subsection{Four orbital tight binding model in $C_3$ format}
In general, the three matrices $D/P^{xy}_{l}$ can be interconnected through the $C_3$ rotational symmetry. Specifically, we have:
\begin{align}
D/P^{xy}_{3} &= C_3^{\dagger} D/P^{xy}_{1} C_3, \\
D/P^{xy}_{2} &= C_3^{\dagger} D/P^{xy}_{3} C_3.
\end{align}
The explicit forms of these matrices are provided below:
\begin{align}
	D_{1}^z=\left(\begin{array}{cc}
		d_{2}^z & -d_{1}^z \\
		d_{1}^z & d_{2}^z
	\end{array}\right)
\end{align}
\begin{align}
	D_{1}^{xy}=\left(\begin{array}{cc}
		d_{\sigma} & d_{1}+d_{2} \\
		d_{1}-d_{2} & d_{\pi}
	\end{array}\right)
\end{align}
\begin{align}
	4D_{2}^{xy}=\left(\begin{array}{cc}
		d_{\sigma}+3 d_{\pi} - 2 \sqrt3 d_1 & \sqrt{3} d_{\sigma}- \sqrt{3} d_{\pi}  -2 d_1+4 d_2)\\
		\sqrt{3} d_{\sigma}- \sqrt{3} d_{\pi} -2 d_1 - 4d_2 &    3 D_{\sigma}+d_{\pi} + 2\sqrt3 d_1
	\end{array}\right)
\end{align}
\begin{align}
	4D_{3}^2=\left(\begin{array}{cc}
		d_{\sigma}+3 d_{\pi}  +  2 \sqrt3 d_1 & -\sqrt{3} d_{\sigma}  + \sqrt{3} d_{\pi} - 2d_1+4d_2 \\
		-\sqrt{3}  d_{\sigma} + \sqrt{3} d_{\pi} - 2 d_1-4 d_2&  3 d_{\sigma}+d_{\pi} - 2\sqrt3 d_1
	\end{array}\right).
\end{align}
\begin{align}
	P_{1}^z=\left(\begin{array}{cc}
		p_{2}^z & -p_{1}^z \\
		p_{1}^z & p_{2}^z
	\end{array}\right)
\end{align}
\begin{align}
	P_{1}^{xy}=\left(\begin{array}{cc}
		p_{\sigma} & p_{1}+p_{2} \\
		p_{1}-p_{2} & d_{\pi}
	\end{array}\right)
\end{align}
\begin{align}
	4P_{2}^{xy}=\left(\begin{array}{cc}
		p_{\sigma}+3 p_{\pi} - 2 \sqrt3 p_1 & \sqrt{3} p_{\sigma}- \sqrt{3} p_{\pi}  -2 p_1+4 p_2)\\
		\sqrt{3} p_{\sigma}- \sqrt{3} p_{\pi} -2 p_1 - 4p_2 &    3 p_{\sigma}+p_{\pi} + 2\sqrt3 p_1
	\end{array}\right)
\end{align}
\begin{align}
	4P_{3}^2=\left(\begin{array}{cc}
		p_{\sigma}+3 p_{\pi}  +  2 \sqrt3 p_1 & -\sqrt{3} p_{\sigma}  + \sqrt{3} p_{\pi} - 2p_1+4p_2 \\
		-\sqrt{3}  p_{\sigma} + \sqrt{3} p_{\pi} - 2 p_1-4 p_2&  3 p_{\sigma}+p_{\pi} - 2\sqrt3 p_1
	\end{array}\right).
\end{align}
Here, the parameters are extracted as: $e_p$ = -312.98 meV, $p_1^z$ = 1.57 meV, $p_2^z$ = -68.23 meV, $p_{\sigma}$ = -4.21 meV, $p_{\pi}$ = 1.70 meV, $p_1$ = -1.35 meV, $p_2$ = 2.53 meV, $e_d$ = -109.51 meV, $d_1^z$ = -3.78 meV, $d_2^z$ =  -8.59 meV, $d_{\sigma}$ = 2.47 meV, $d_{\pi}$ = -7.58 meV, $d_1$ = -4.94 meV, and $d_2$ = -0.95 meV.

\subsection{Super-exchange from perturbation theory}
\begin{figure}[!h]
	\includegraphics[width= 0.8\columnwidth]{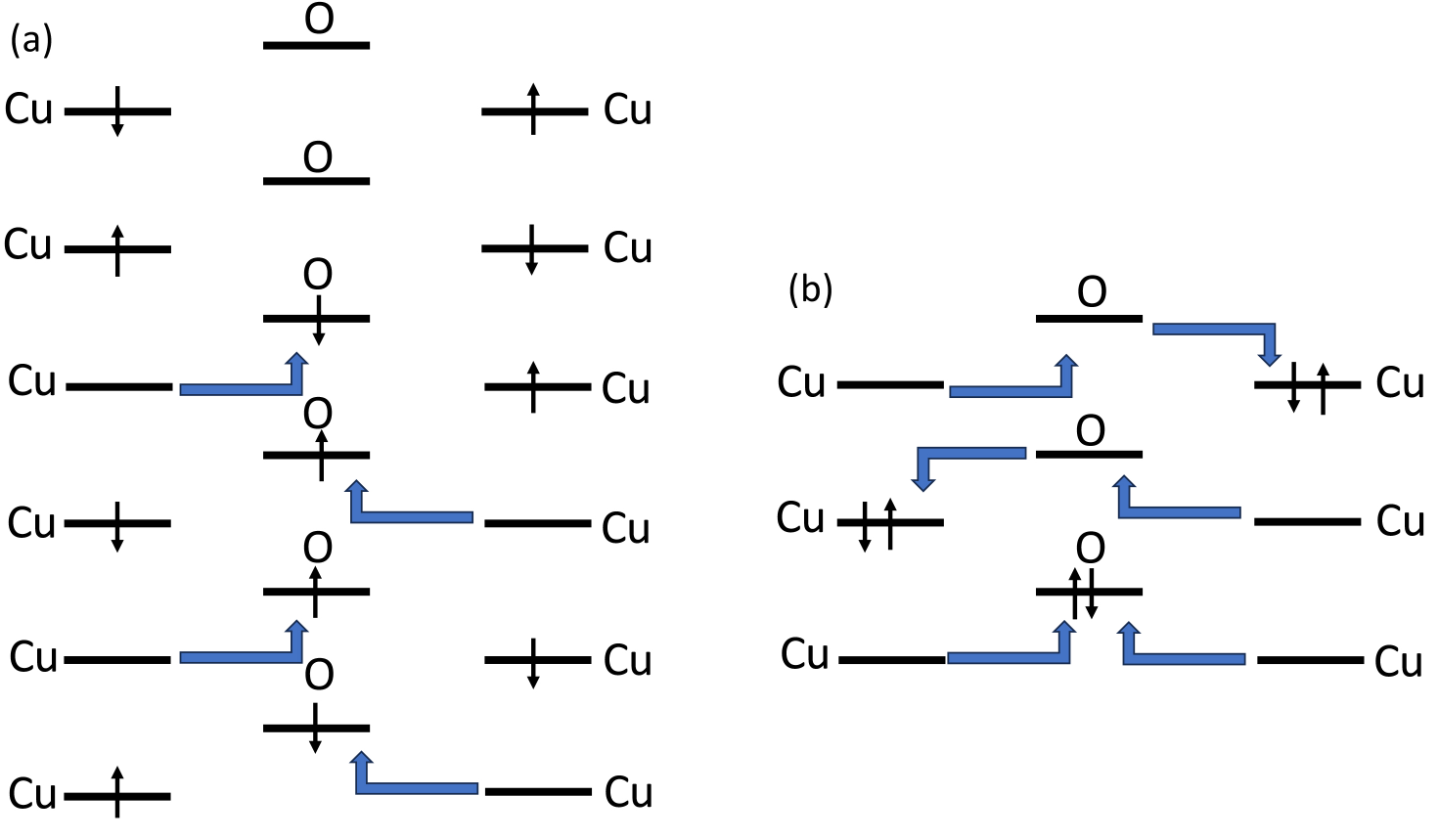}
	\caption{Basis states for the description of superexchange are grouped in the order they appear in the matrix form. (a) One electron transfer states;
 (b) Two electrons transfer states. All spin transfer processes are shown by blue arrows.}
 \label{super_exchange}
\end{figure}

In this section we derive super-exchange coupling with the help of perturbation theory for the Hubbard model Hamiltonian formulated in the main text \eqref{Hubbard_ham}. In order to make the model simple, we will restrict ourselves to a
two-bands model of only one orbital for a single Cu ion ((described by operators $c_{i\sigma}$, $c^\dagger_{i\sigma}$)) and one for O atom ($c_{p\sigma}$, $c^\dagger_{p\sigma}$) instead of a complete four-bands model suggested in the main text. However, as we discuss below, this simplified model proves to be sufficient for a super-exchange order of magnitude estimate. In what was said above $i=1,2$ enumerates Cu atoms and $\sigma$ is spin $1/2$ index. The model two-bands Hamiltonian is then given by
\begin{align}
    H=\sum_\sigma\left(\varepsilon_d \sum_i n_{i \sigma}+\varepsilon_p n_{p \sigma}-t_{p d} \sum_i\left(c_{i \sigma}^{\dagger} c_{p \sigma}+c_{p \sigma}^{\dagger} c_{i \sigma}\right)\right)+U_d \sum_i n_{i \uparrow} n_{i \downarrow}+U_p n_{p \uparrow} n_{p \downarrow}
    \label{ham_simplified}
\end{align}
In what follows we assume that $\varepsilon_d+\varepsilon_p=0$ and $\Delta_{pd}=\varepsilon_p-\varepsilon_d$. Then in the sector of initially antiparallel Cu electrons spin basis, the matrix form of the Hamiltonian \eqref{ham_simplified} reads
\begin{align}
   H= \left(\begin{array}{cc|cccc|ccc}
0 & 0 & -t_{p d} & -t_{p d} & 0 & 0 & 0 & 0 & 0 \\
0 & 0 & 0 & 0 & -t_{p d} & -t_{p d} & 0 & 0 & 0 \\
\hline-t_{p d} & 0 & \Delta_{p d} & 0 & 0 & 0 & -t_{p d} & 0 & -t_{p d} \\
-t_{p d} & 0 & 0 & \Delta_{p d} & 0 & 0 & 0 & -t_{p d} & -t_{p d} \\
0 & -t_{p d} & 0 & 0 & \Delta_{p d} & 0 & +t_{p d} & 0 & +t_{p d} \\
0 & -t_{p d} & 0 & 0 & 0 & \Delta_{p d} & 0 & +t_{p d} & +t_{p d} \\
\hline 0 & 0 & -t_{p d} & 0 & +t_{p d} & 0 & U_d & 0 & 0 \\
0 & 0 & 0 & -t_{p d} & 0 & +t_{p d} & 0 & U_d & 0 \\
0 & 0 & -t_{p d} & -t_{p d} & +t_{p d} & +t_{p d} & 0 & 0 & U_p+2\Delta_{p d}
\end{array}\right)
\label{matrixform}
\end{align}
The first 2 basis vectors of \eqref{matrixform} describe antiparallel Cu spin configurations with an empty O orbital, the following 4 vectors stand for all one-electron transfer processes and the last 3 vectors describe two-electron processes leading to double occupation of a single orbital (for a visual represenation of basis vectors please see Fig. \ref{super_exchange}).
%Note that in \eqref{matrixform} we did not consider the sector of initially parallel Cu spins since in this case Pauli principle does not allow for an for the reason that .
Performing perturbation theory approach in powers of $1/U$, for the effective Hamiltonian of interacting antiparallel Cu electrons spins we arrive at
\begin{align}
    H_\mathrm{eff}=-\frac{2t_{pd}^4}{\Delta_{pd}^2}\left(\frac{1}{U_d}+\frac{1}{U_p+2\Delta_{pd}}\right)\begin{pmatrix}
                                1 & -1\\
                                -1 & 1
    \end{pmatrix},
\end{align}
so that super-exchange coupling is given by
\begin{align}
    J=\frac{4t^4_{p d}}{\Delta_{pd}^2}\left(\frac{1}{U_d}+\frac{1}{U_p+2\Delta_{pd}}\right).
    \label{superex}
\end{align}
Let us note that for the complete 4-band model every $t_{pd}$ in \eqref{matrixform} should be replaced by 2x2 matrices from \eqref{transition_matrices}. However, this modification would lead only to a more complex $t$-dependence in \eqref{superex} which should not change the order of magnitude value of $J$.

\end{document}